\documentclass{PoS}

\usepackage{graphicx}

\title{Phase Transitions in Dense and Hot Matter}

\ShortTitle{PT's in Dense and Hot Matter}

\author{\speaker{Veronica Dexheimer}\\
        Department of Physics, Kent State University, Kent OH, USA\\
        E-mail: \email{vdexheim@kent.edu}}

\abstract{In this conference proceeding, I discuss in detail the deconfinement to quark matter that takes place at large densities and/or temperatures. The first-order phase transition that is assumed to appear beyond a critical point gives rise to mixtures of phases when more than one globally conserved quantity (such as charge fraction) is imposed. The modifications caused by these mixtures of phases in the QCD phase diagram can have consequences on signals of the existence of quark matter expected to be created in heavy-ion collisions, as well as supernova explosions and neutron-star mergers.}

\FullConference{Critical Point and Onset of Deconfinement - CPOD2017\\
		7-11 August, 2017\\
		The Wang Center, Stony Brook University, Stony Brook, NY}

\begin{document}

At low and zero chemical potentials, it has been known for quite some time that the quark deconfinement in isospin-symmetric matter is a smooth crossover \cite{Aoki:2006we}. At larger chemical potentials, on the other hand, there are only hints that there is a critical point, beyond which quark deconfinement takes place in the form of a first-order phase transition. The relevant data include flow measurements near mid-rapidity recorded by the STAR detector at the Relativistic Heavy Ion Collider (RHIC) \cite{Adamczyk:2014ipa} and measurements
of the cumulants (up to fourth order) of event-by-event multiplicity distributions from the first phase of the beam energy scan program at RHIC \cite{Luo:2017faz}.

To study the QCD phase diagram and the characteristics of the chiral symmetry restoration and quark deconfinement phase transitions, I employ an effective relativistic model, the Chiral Mean Field (CMF) model \cite{Dexheimer:2008ax,Dexheimer:2009hi,Dexheimer:2015qha}. It is an extended non-linear realization of the SU(3) sigma model, which uses pseudo-scalar mesons as parameters of the chiral transformation \cite{Papazoglou:1998vr}. It includes the baryon octet, leptons, and quarks as degrees of freedom and was fitted to reproduce nuclear, lattice QCD, heavy-ion collision, and astrophysical constraints. In this model, the degrees of freedom change from hadrons to quarks through a contribution of the field $\Phi$ (named in analogy to the Polyakov loop) to their effective masses, which are generated by the interaction with the medium:
\begin{eqnarray}
M_{i}^*&=&g_{i\sigma}\sigma+g_{i\delta}\tau_3\delta+g_{i\zeta}\zeta+M_{0_i}\,,
\end{eqnarray}
where $M_{0_i}$ are small explicit mass terms. The values for the model couplings, together with the bare mass values and $\Phi$ contributions can be found in Ref.~\cite{Dexheimer:2009hi}.

A potential for $\Phi$, which acts as an order parameter for deconfinement,
\begin{eqnarray}
U=(a_0T^4+a_1\mu^4+a_2T^2\mu^2)\Phi^2+a_3T_0^4\log{(1-6\Phi^2+8\Phi^3-3\Phi^4)}\,,
\end{eqnarray}
is present at all chemical potentials and temperatures and generates the first-order phase transition (quark deconfinement) coexistence lines seen in Fig.~1. The circles mark critical points, beyond which the deconfinement and chiral transitions become smooth crossovers, and individual phases cannot be further identified. The shaded regions in the figure exemplify some of the different regimes that can be described within the CMF model (that can be applied to the entire $\mu_B-T$ plane) and how these different regimes, involving laboratory experiments and astrophysical objects/events, can overlap. The lower black line shows that for isospin-symmetric matter, the nuclear liquid-gas phase transition is reproduced (within the CMF model) at low chemical potentials and temperatures. For large chemical potentials and temperatures, the black and pink lines stand for the deconfinement lines for isospin-symmetric matter with zero net strangeness and charge neutral and chemically equilibrated matter, respectively (again calculated within the CMF model). It is important to point out that the deconfinement critical points were obtained under the assumption that, for finite temperature, there are quarks in the hadronic phase and hadrons in the quark phase and each phase is defined by the value of the order parameter for deconfinement $\Phi$. Its values become more similar across the phase transition as the temperature increases and, finally coincide at the critical point.

Phase transitions in systems with macroscopic phases that possess more than one globally conserved charge are of non-congruent type (see Refs.~\cite{Iosilevskiy:2010qr,Hempel:2013tfa,Hempel:2015eoj} and references therein for details). In these kind of phase transitions, the local concentration of the charge associated with the conserved quantity varies in the two coexisting macroscopic phases and the associated chemical potential is the same in both phases. Here, two examples are shown concerning the deconfinement phase transition to quark matter:

\begin{figure}[t!]
\begin{center}
\includegraphics[width=0.8\textwidth]{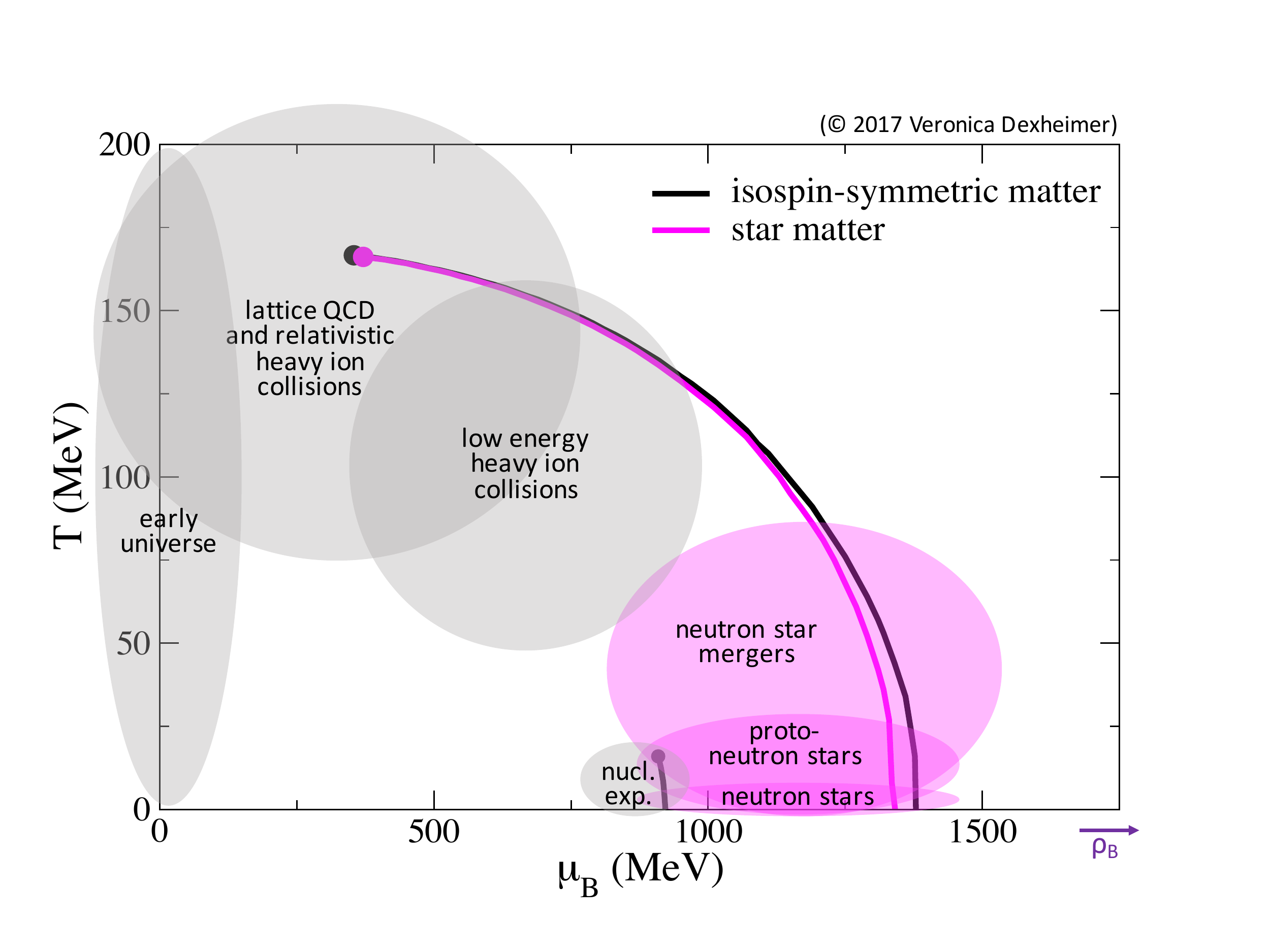}
\caption{(Color online) QCD phase diagram for isospin-symmetric matter (with zero net strangeness) and neutron-star matter (charge neutral and in chemical equilibrium) calculated using the CMF model.}
\end{center}
\end{figure}

\begin{itemize}
\item with fixed baryon number and charge fraction $Y_c=0.5$, $Y_c=0.3$ (with zero net strangeness)
\item with fixed baryon number and zero charge (in chemical equilibrium)
\end{itemize}

The first case in which baryon number and fixed (but not zero) charge fraction are conserved is going to be referred to as HI, as this kind of matter is created in heavy-ion collisions. Charge fraction is defined as the total baryonic charge over total baryon number
\begin{eqnarray}
Y_q=\frac Q B=\frac {\sum_i Q_{e_i} n_i}{\sum_i  {Q_B}_i n_i}\,,
\label{yc}
\end{eqnarray}
where ${Q_e}_i$ is the electric charge, ${Q_B}_i$ is the baryon number, and $n_i$ the number density of each baryon or quark. Note that $\sum {Q_B}_i n_i$ is not the same as the baryon number density ${n_B}$, as the latter comes from the derivative of the pressure with respect to the baryon chemical potential and, therefore, also contains a contribution from the potential $U$ for $\Phi$ (when quarks are present).

\begin{figure}[t!]
\begin{center}
\includegraphics[trim={.95cm 0cm 0cm 0cm},width=9.cm]{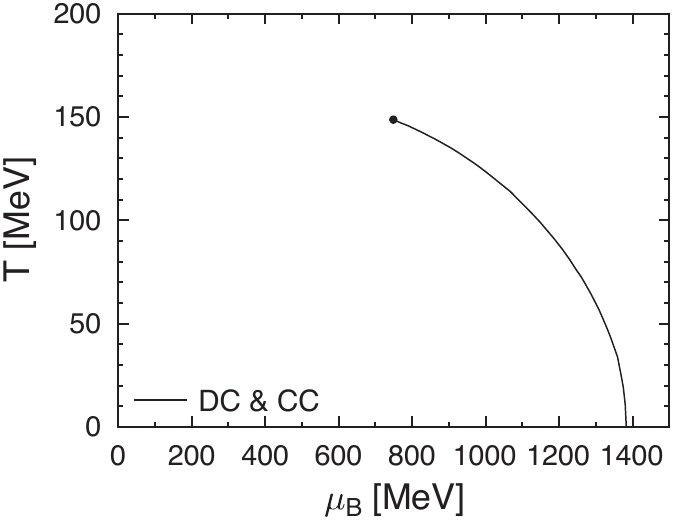}
\caption{(Color online) Phase diagram for isospin-symmetric (HIS) matter with charge fraction $Y_q=0.5$ showing temperature vs. baryon chemical potential. DC stands for  deconfinement curve and CC for confinement curve.}
\end{center}
\end{figure}

Although the charge fraction in Au-Au or Pb-Pb collisions is $\sim$  0.4, for very energetic collisions this does not matter, as matter created behind the collision in the fireball has no net baryon number density. For lower energy collisions, on the other hand, baryonic matter experiences "stopping" and the charge fraction involved is important.  In this work, two charge fractions are studied and compared, $Y_c=0.5$ and $Y_c=0.3$.

Since strangeness is set to zero when describing matter generated in heavy-ion collisions, charge fraction $Y_c=0.5$ is a synonym to isospin-symmetric matter $Y_{\rm{iso}}=0$. This can be easily verified starting from the relation between isospin projection, electric charge, baryon number, and strangeness quantum numbers of each baryon or quark
\begin{eqnarray}
I_i=Q_{e_i}-\frac 1 2 Q_{B_i}+ \frac 1 2 S_i\,,
\label{yc}
\end{eqnarray}
which, summed over all baryons and quarks and divided by the baryons number, gives
\begin{eqnarray}
\frac{I}{B}=\frac{Q_e}{B}-\frac 1 2 \frac{B}{B}+ \frac 1 2 \frac{S}{B}\,,
\label{yc}
\end{eqnarray}
\begin{eqnarray}
Y_{\rm{iso}}=Y_q-\frac 1 2 + \frac 1 2 Y_s\,,
\label{yc}
\end{eqnarray}

For the half-charged matter $Y_c=0.5$ (isospin-symmetric case HIS), the chemical potential associated with the conserved quantity charge (or isospin), $\mu_q$, is necessarily zero so the proton and neutron have the same chemical potential. This results in an azeotropic behavior \cite{Muller:1995ji} and a necessarily congruent phase transition. Figure~2 shows the respective phase diagram using the CMF model. In this case, there is no extended phase coexistence region and the deconfinement and confinement curves coincide. Note that, if temperature were plotted against pressure instead, the slope of the coexistence line would still be negative, which is the opposite behavior from the liquid-gas type of phase transition. For details concerning this issue, see Refs.~\cite{Iosilevskiy:2014qha,Hempel:2013tfa,Dexheimer:2017ecc}.

Fig.~3 illustrates the non-congruent features that take place in the case of a fixed charge fraction $Y_q=0.3$ (isospin-asymmetric case HIAS). The phase coexistence occupies an extended region of the diagram going from the confinement curve until the deconfinement curve. Inside, the charged chemical potential (which is the same in each phase) changes continuously, as has already been discussed in Refs.~\cite{Muller:1997tm,Shao:2011fk,Sissakian:2006dn}. Nevertheless, the area within this region becomes vanishingly small for large temperatures. Such an effect is not observed for the nuclear liquid-gas phase transition region that extends to much lower temperatures \cite{Barranco:1980zz,Muller:1995ji}. In Fig.~3, there is also a curve calculated for a forced congruent phase transition, in which the charge fraction is forced to be the same in both phases. In this case, shown only for comparison, the charge fractions are conserved locally in each phase and each phase has a distinct charged chemical potential. Finally, note that the x-axis in Fig.~3 is $\tilde{\mu}_B$, which is the chemical potential that is the same in both phases (${\mu}_B$ is not). It is defined as $\tilde{\mu}_B={\mu}_B+Y_q\mu_q$ (see Ref.~\cite{Hempel:2009vp,Hempel:2013tfa,Dexheimer:2017nse} for details).

In neutron stars, the conserved quantities are baryon number and zero charge. In addition, matter is in beta equilibrium and contains net strangeness. In the astrophysics community, non-congruent phase transitions are referred to as Gibbs constructions \cite{Glendenning:1992vb} and congruent phase transitions (physically forced by a possibly large surface tension between phases) are called Maxwell constructions. Figure~4 illustrates those two scenarios together with an example trajectory for a cold neutron star at zero temperature. The most relevant non-congruent feature that can take place in neutron stars is the appearance of a coexistence of charged microfragments of hadrons and quarks that extends over a large portion of the star's radius (see for example Ref.~\cite{Dexheimer:2009hi}). Note that in Fig.~1 the neutron-star matter phase transition is congruent because of the assumption of local charge neutrality.

\begin{figure}[t!]
\begin{center}
\includegraphics[trim={.95cm 0cm 0cm 0cm},width=9.cm]{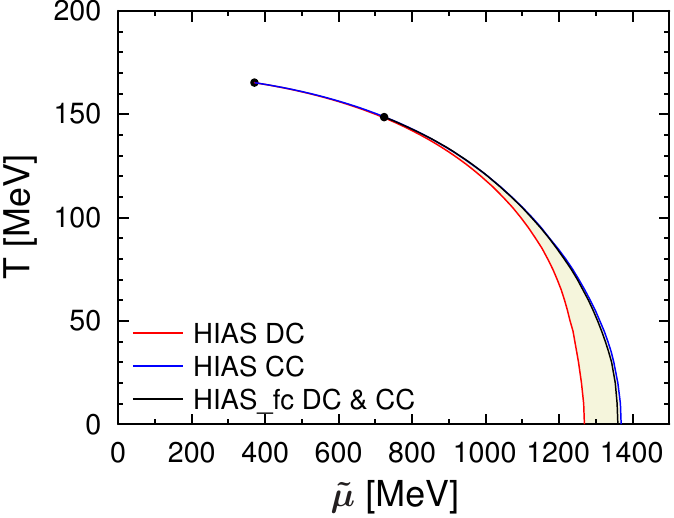}
\caption{(Color online) Phase diagram for isospin-asymmetric (HIAS) matter with charge fraction $Y_q=0.3$ showing temperature vs. modified baryon chemical potential. DC stands for  deconfinement curve and CC for confinement curve. $f_c$ illustrates the non-physical forced congruent case.}
\end{center}
\end{figure}

\begin{figure}[t!]
\begin{center}
\includegraphics[trim={.95cm 0cm 0cm 0cm},width=9.cm]{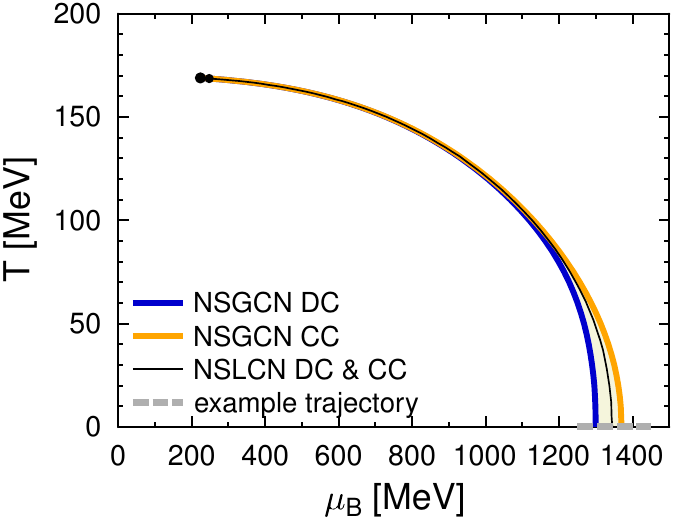}
\caption{(Color online) Phase diagram for neutron-star (NS) matter showing temperature vs. baryon chemical potential assuming local (NSLCN) and global charge neutrality (NSGCN). DC stands for  deconfinement curve and CC for confinement curve.}
\end{center}
\end{figure}

Summarizing, in order to have a complete description of dense and hot matter, different degrees of freedom have to be considered. These include necessarily nucleons, hyperons, and particularly when describing hot matter, quarks. Note that, as shown in Fig.~1, hot matter not only applies to matter created in heavy-ion collisions, but also matter created inside the core of supernova explosions and generated during different stages of neutron star mergers. The CMF has been calibrated and is suitable to describe the entire QCD phase diagram, including the description of critical points. It is in agreement with zero temperature nuclear physics, astrophysics, heavy-ion collisions, and lattice QCD. The model is consistent with perturbative QCD at zero temperature (Fig.~1 of Ref.~\cite{Dexheimer:2017pom}) and finite temperature \cite{Jake}. 

In this work, the CMF model was used to study the thermodynamics of the QCD phase diagram and draw comparisons between possible mixtures of phases generated by assuming different constraints. Although these results are quantitatively model dependent, the qualitative statements are not. Therefore, 
the appearance of mixtures of phases should be taken in to account when looking for signals for deconfinement in the lab and in the sky.

The author would like to thank Matthias Hempel and Igor Iosilevskiy for helpful discussions. The author acknowledges the support from NewCompStar COST Action MP1304.

\bibliographystyle{JHEP}
\bibliography{skeleton}

\end{document}